\documentclass[conference]{IEEEtran}
\IEEEoverridecommandlockouts
\usepackage{cite}
\usepackage{amsmath,amssymb,amsfonts}
\usepackage{algorithmic}
\usepackage{graphicx}
\usepackage{booktabs}
\usepackage{adjustbox}
\usepackage[compatibility=false]{caption}
\usepackage{subcaption}
\usepackage{textcomp}
\usepackage{xcolor}
\usepackage{comment}
\usepackage{enumitem}
\usepackage{glossaries}
\usepackage{microtype}
\usepackage{hyperref}
\usepackage{xurl}
\newacronym{bs}{BS}{base station}
\newacronym{ue}{UE}{user equipment}
\newacronym{nr}{NR}{new radio}
\newacronym{mimo}{MIMO}{multiple-input multiple-output}
\newacronym{mt}{MT}{mobile terminal}
\newacronym{toa}{ToA}{time of arrival}
\newacronym{tdoa}{TDoA}{time difference of arrival}
\newacronym{rtt}{RTT}{round-trip time}
\newacronym{rss}{RSS}{received signal strength}
\newacronym{aoa}{AoA}{angle of arrival}
\newacronym{fr1}{FR1}{frequency range 1}
\newacronym{cp}{CP}{carrier phase}
\newacronym{rtk}{RTK}{real-time kinematic}
\newacronym{ppp}{PPP}{precise point positioning}
\newacronym{gnss}{GNSS}{global navigation satellite system}
\newacronym{los}{LoS}{line of sight}
\newacronym{fr2}{FR2}{frequency range 2}
\newacronym{crlb}{CRLB}{Cramér-Rao lower bound}
\newacronym{inf}{InF}{indoor factory}
\newacronym{uma}{UMa}{urban macro}
\newacronym{rmse}{RMSE}{root mean square error}
\newacronym{prs}{PRS}{positioning reference signal}
\newacronym{awgn}{AWGN}{additive white Gaussian noise}
\newacronym{psd}{PSD}{power spectral density}
\newacronym{snr}{SNR}{signal-to-noise ratio}
\newacronym{pdf}{PDF}{probability density function}
\newacronym{ml}{ML}{maximum likelihood}
\newacronym{ecdf}{ECDF}{empirical cumulative distribution function}
\newacronym{wls}{WLS}{weighted least-squares}

\usepackage[top=0.765in,bottom=1.04in,left=0.625in,right=0.625in]{geometry}

\def\BibTeX{{\rm B\kern-.05em{\sc i\kern-.025em b}\kern-.08em
    T\kern-.1667em\lower.7ex\hbox{E}\kern-.125emX}}
\begin{document}

\title{Accuracy of Joint Time-Based and Carrier-Phase Positioning in 5G Networks under Correlated Measurement Errors}

{\author{
\IEEEauthorblockN{
Nahidul Islam\IEEEauthorrefmark{1},
Mohammad Razzaghpour\IEEEauthorrefmark{1}\IEEEauthorrefmark{2},
Marwan Hammouda\IEEEauthorrefmark{1},
Carsten Bockelmann\IEEEauthorrefmark{1},
and Armin Dekorsy\IEEEauthorrefmark{1}\IEEEauthorrefmark{2}
}
\IEEEauthorblockA{\IEEEauthorrefmark{1} Dept.\ of Communications Engineering, University of Bremen, Germany}
\IEEEauthorblockA{\IEEEauthorrefmark{2} Gauss-Olbers Space Technology Transfer Center, University of Bremen, Germany}
\IEEEauthorblockA{Emails: nahidul@uni-bremen.de, \{razzaghpour, hammouda, bockelmann, dekorsy\}@ant.uni-bremen.de}

\thanks{This work is supported by the Federal Ministry for Economic Affairs and Climate Action, based on a decision of the German Bundestag, and has been submitted to the IEEE for possible publication. Copyright may be transferred without notice, after which this version may no longer be accessible.}
}

\maketitle

\begin{abstract}

High-accuracy positioning is critical for emerging applications such as autonomous driving, industrial automation, augmented reality, and smart cities. 3GPP Release 18 introduced carrier-phase~(CP) positioning for 5G that offers superior accuracy compared to conventional time-based methods such as time of arrival~(ToA). However, CP-based positioning requires resolving the integer phase ambiguity, which refers to the unknown number of full-wavelength cycles completed during signal propagation. Joint processing of ToA and CP can mitigate this integer ambiguity by narrowing down the search space of possible integers, particularly for short wavelengths. This paper investigates the performance of a positioning method that integrates ToA and CP measurements. As a main contribution, the analysis explicitly accounts for the error correlation between ToA and CP measurements. Furthermore, the study analyzes the impact of key 5G system parameters on positioning accuracy using this correlation-aware joint method in both factory and urban environments, where many 5G positioning applications are expected to emerge. The results highlight that exploiting this correlation can further improve positioning performance by approximately $7$ percent. Moreover, the findings of this study provide insight into how 5G system parameters can be tuned to achieve centimeter-level accuracy under favorable conditions.

\end{abstract}


\section{Introduction}

As wireless technology evolves from 5G to 5G-Advanced and eventually to 6G, positioning has emerged as a key capability \cite{10437902}.
By leveraging wider bandwidths, higher carrier frequencies, \gls{mimo}, and dense network infrastructure, 5G can achieve positioning accuracy that was previously unattainable in earlier generations of cellular networks.
This improved accuracy is essential for enabling a wide range of emerging applications, from autonomous driving and industrial robotics to augmented reality and smart cities.

Conventional radio positioning techniques determine the location of a \gls{mt} through \gls{toa}, \gls{rss}, and \gls{aoa} measurements. The accuracy of these methods is limited by factors such as measurement precision and resolution, signal processing techniques, and the radio propagation environment. In 5G, traditional time-based methods can achieve sub-meter accuracy under favorable conditions, e.g., when the maximum available bandwidth is used \cite{10332505}.

To further improve positioning accuracy, 3GPP Release 18 has recently introduced \gls{cp} positioning techniques for 5G \cite{keating2021evolution, 3gpp38859}.
This idea builds on the well-established use of \gls{cp}-based methods in \gls{gnss}, such as \gls{rtk} and \gls{ppp}, which can achieve centimeter-level accuracy when \gls{gnss} satellite signals are available.

Generally, \gls{cp}-based methods suffer from the so-called integer phase ambiguity, defined as the unknown number of whole carrier wavelengths traversed by the radio wave during propagation from transmitter to receiver, since the phase repeats itself every complete cycle, i.e., every $2\pi$. The main challenge in resolving this phase ambiguity is the large search space of possible integers, especially for mmWave frequencies.

In \cite{10437192}, integer ambiguity is resolved using a multi-frequency carrier method. Alternatively, when multi-frequency carrier-phase measurements are unavailable, the search space for the integer ambiguities can be reduced by combining time-based measurements with a \gls{cp}-based approach \cite{10475845}.
Regarding techniques that integrate \gls{toa} and \gls{cp} measurements, \cite{10437902} employs analytical tools such as the \gls{crlb} to determine their theoretical positioning accuracy limits, thereby providing a benchmark for evaluating practical implementations.
However, it does not consider the correlations between the measurement errors of \gls{toa} and \gls{cp}, which may exist in practical setups.
This correlation arises because \gls{toa} and \gls{cp} are derived from the same received signal and are affected by common error sources, such as multipath effects, clock synchronization mismatch between the \glspl{bs} and the \gls{mt}, and thermal noise. However, the impact of correlated \gls{toa} and \gls{cp} measurement errors has received limited attention in the existing literature. 

In this paper, we propose and evaluate a joint positioning method that integrates \gls{toa} and \gls{cp} measurements while explicitly accounting for correlated measurement errors. Furthermore, using this correlation-aware joint method, we investigate the impact of key 5G parameters on positioning accuracy, including carrier frequency, bandwidth, transmit power, antenna configurations, and the number of involved \glspl{bs}. 

%The rest of this paper is organized as follows: Section II introduces the joint positioning method, including the system model, the observation model, and the position estimation algorithm. Section III describes the simulation setup, consisting of the path loss model, network configuration, and accuracy metrics. Section IV presents the results and discusses the impact of network parameters and measurement-error correlation on positioning accuracy. Finally, Section V concludes the paper.

The rest of the paper is organized as follows: Section~II introduces the correlation-aware joint positioning method, Section~III outlines the simulation setup, Section~IV presents and discusses the results, and Section~V concludes the paper.

\section{Correlation-Aware Joint Positioning Method} 

\subsection{System and Received Signal Model}
Consider a 5G network, as shown in Fig.~\ref{fig:positioning_problem}, comprising an \gls{mt} located at an unknown position $\mathbf{x} = \left[x \ y\right]^\mathsf{T}$ and $N \ge 3$ \glspl{bs} with known positions $\mathbf{x}_i = \left[x_i \ y_i\right]^\mathsf{T}$. 
This study considers downlink positioning only.
Let $s(t)$ denote the transmitted downlink \gls{prs} from the $i$-th \gls{bs}. 
Assuming a line-of-sight wireless channel, the baseband-equivalent received signal at the \gls{mt} from the $i$-th \gls{bs} is given by
\begin{equation}
        y_i(t) = a_i e^{j\phi_i} s(t-\tau_i) + w_i(t),
\end{equation}
where $a_i$ is the attenuation, $\tau_i$ is the propagation delay (\gls{toa}), $\phi_i$ is the carrier phase, and $w_i(t)$ is the zero-mean \gls{awgn} with variance $\sigma_w^2 = N_0B$, where $N_0$ is the \gls{psd}  and $B$ is the bandwidth.
The attenuation $a_i$ captures the combined effects of path loss and antenna gains, as follows
\begin{equation}\label{amp}
    a_i = \sqrt{P_{\mathrm{BS}_i} G_{\mathrm{BS}_i} G_{\mathrm{MT}} \left (\frac{\lambda}{4\pi d_0}\right )^2 \left (\frac{d_0}{d_i}\right )^{\gamma}},
\end{equation}
where $P_{\mathrm{BS}_i}$ is the transmit power of \gls{bs}, $G_{\mathrm{BS}_i}$ and $G_{\mathrm{MT}}$ are antenna gains of \gls{bs} and \gls{mt} respectively, $\lambda$ is the carrier wavelength, $\gamma$ is the path loss exponent, $d_0$ is the reference distance, and $d_i = {\|\mathbf{x} - \mathbf{x}_i\|}_2$ is the distance between the \gls{mt} and the $i$-th \gls{bs}. The delay and the phase are geometrically related by
\begin{equation}
    \tau_i = \frac{d_i}{c} \quad \text{and} \quad \phi_{i} = \frac{2\pi d_i}{\lambda}.
\end{equation}

\subsection{Joint Observation Model} 
From the received signal,  \gls{toa}- and \gls{cp}-based range measurements are obtained as 
\begin{equation}
    r_{\tau_i} = c\hat{\tau}_i = d_i + n_{\tau_i},
\end{equation}
\begin{equation}
    r_{\phi_i} = \hat{z}_i\lambda+\frac{\lambda}{2\pi}\hat{\phi}_i = d_i + n_{\phi_i},
\end{equation}
where $n_{\tau_i}$ and $n_{\phi_i}$ are assumed to be zero-mean Gaussian measurement errors, considering perfectly synchronized \glspl{bs} and negligible calibration offsets. The integer phase ambiguity estimate $\hat{z}_i$ is obtained using the traditional least-squares ambiguity decorrelation adjustment~(LAMBDA) algorithm \cite{teunissen1997least} after calculating the initial estimate by
\begin{equation}
    \hat{z}_i^{\mathrm{init}} = \left \lfloor \frac{c \hat{\tau}_i}{\lambda} \right \rfloor.
\end{equation}
The standard deviations of $n_{\tau_i}$ and $n_{\phi_i}$ are given by \cite{pahlavan2022indoor, kay1993fundamentals, 10012896}
\begin{equation}\label{toaerror}
\sigma_{\tau_i} \propto \frac{1}{B\sqrt{\mathrm{SNR}_i}}, \quad 
\sigma_{\phi_i} \propto \frac{1}{f_c\sqrt{\mathrm{SNR}_i}},
\end{equation}
respectively, where $\mathrm{SNR}_i$ is the \gls{snr} and $f_c$ is the carrier frequency. Stacking observations from all $N$ \glspl{bs}, the joint observation model can be expressed as
\begin{equation}\label{rtoacp}
\mathbf{r} = \boldsymbol{\mu}(\mathbf{x}) + \mathbf{n},
\end{equation}
where $\mathbf{r} = \left[\mathbf{r}^\mathsf{T}_{\tau} \ \mathbf{r}^\mathsf{T}_{\phi}\right]^\mathsf{T} \in \mathbb{R}^{2N}$ and $\mathbf{n} = \left[\mathbf{n}^\mathsf{T}_{\tau} \ \mathbf{n}^\mathsf{T}_{\phi}\right]^\mathsf{T} \in \mathbb{R}^{2N}$. The mean vector $\boldsymbol{\mu}(\mathbf{x})$ contains the true geometric distances between the \gls{mt} and the \glspl{bs}. 

\begin{figure}
    \centering
    \includegraphics[width=0.825\linewidth]{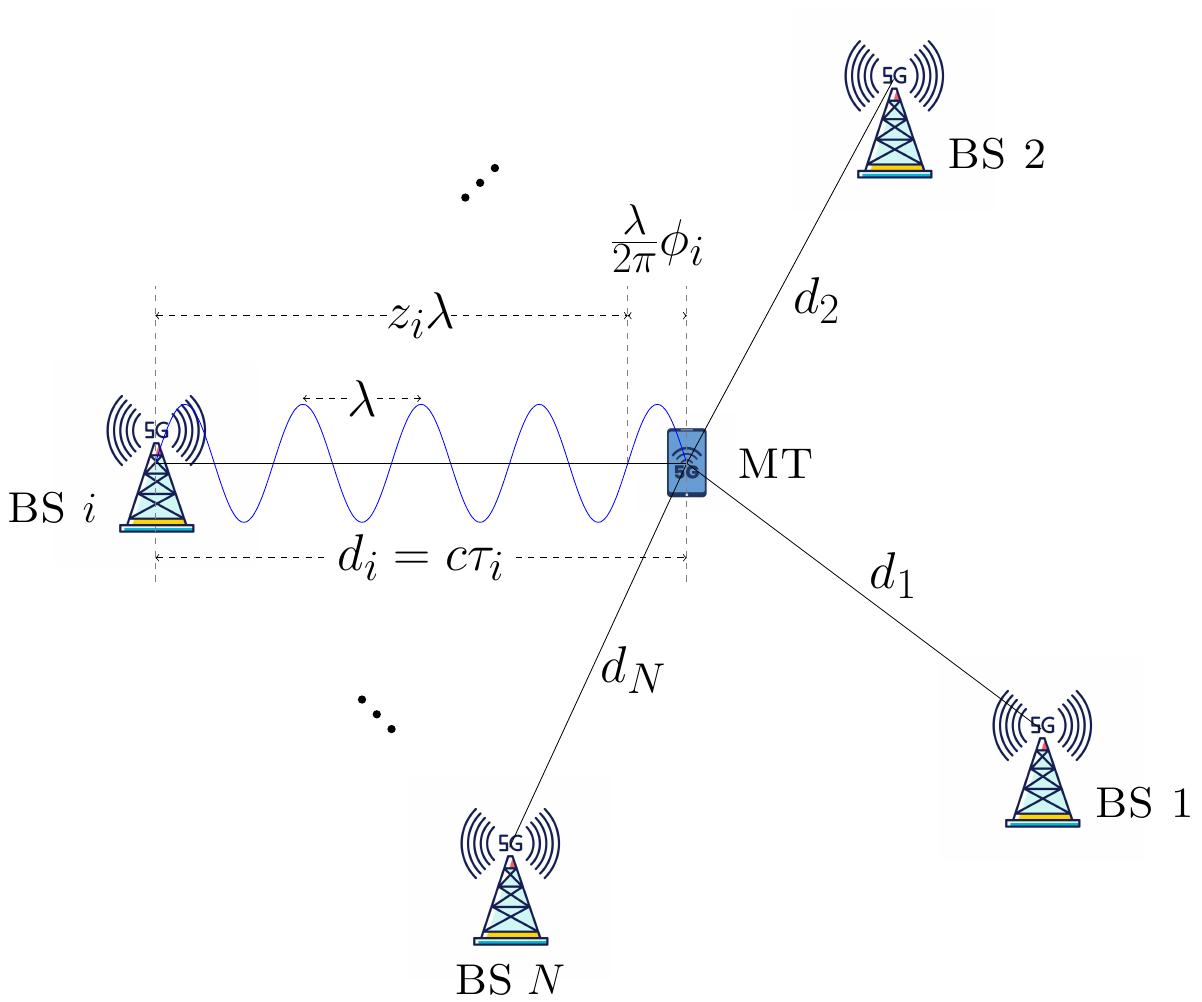}
    \caption{ToA- and CP-based positioning measurements in 5G networks.}
    \label{fig:positioning_problem}
\end{figure}

\subsection{Proposed Error Correlation Model}

We propose an error correlation model to explicitly account for the statistical dependence between ToA and CP measurement errors. To this end, the covariance matrix of the error vector $\mathbf{n}$ includes non-zero cross-covariance terms and is given by
\begin{equation}
    \mathbf{C} = 
    \begin{bmatrix}
    \mathbf{R}_\tau & \mathbf{R}_{\tau\phi} \\
    \mathbf{R}_{\tau\phi}^\mathsf{T} & \mathbf{R}_\phi
    \end{bmatrix} \in \mathbb{R}^{2N\times2N}.
\end{equation}
Since the measurement errors are assumed to be uncorrelated across different \glspl{bs}, the covariance matrices  $\mathbf{R}_\tau$ and $\mathbf{R}_\phi$ can be written as
\begin{equation}\label{error}
    \mathbf{R}_\tau = \mathrm{diag}(\sigma_{\tau_1}^2, \dots, \sigma_{\tau_N}^2) \in \mathbb{R}^{N \times N}
\end{equation} 
and
\begin{equation}
    \mathbf{R}_\phi = \mathrm{diag}(\sigma_{\phi_1}^2, \dots, \sigma_{\phi_N}^2) \in \mathbb{R}^{N \times N}.
\end{equation}
As expressed in~\eqref{toaerror}, the individual error variances depend on the bandwidth, carrier frequency, and the received \gls{snr}. The cross-covariance matrix is given by
\begin{equation}\label{Rtoacp}
    \mathbf{R}_{\tau\phi} = \mathrm{diag}(\rho_{\tau_1\phi_1} \sigma_{\tau_1} \sigma_{\phi_1}, \dots, \rho_{\tau_N\phi_N} \sigma_{\tau_N} \sigma_{\phi_N}) \in \mathbb{R}^{N\times N},
\end{equation}
where $\rho_{\tau_i\phi_i}$ is the correlation coefficient between ToA and CP measurement errors for the $i$-th \gls{bs}, with $|\rho_{\tau_i\phi_i}| < 1$. 
This formulation incorporates error correlations through non-zero cross-covariance terms, thereby enabling the estimator to exploit the resulting information gains.

\subsection{Position Estimation Model}
The conditional \gls{pdf} of $\mathbf{r}$ given $\mathbf{x}$ is expressed as
\begin{equation}\label{eq:pdf}
\begin{split}
    & p(\mathbf{r} \mid \mathbf{x}) \\& = \frac{1}{(2\pi)^{N} \sqrt{\lvert \mathbf{C} \rvert}} 
\exp\left(
-\frac{1}{2}
\left(\mathbf{r} - \boldsymbol{\mu}(\mathbf{x})\right)^{\mathsf{T}}
\mathbf{C}^{-1}
\left(\mathbf{r} - \boldsymbol{\mu}(\mathbf{x})\right)
\right),
\end{split}
\end{equation}
where $\lvert \mathbf{C} \rvert$ denotes the determinant of the covariance matrix of the error vector $\mathbf{n}$. Since the measurement errors are assumed to be Gaussian, the estimate of the \gls{mt} position can be obtained using the \gls{ml} estimator by maximizing the \gls{pdf}, i.e.,
\begin{equation}
\label{eq:ml_estimator}
\hat{\mathbf{x}} 
= \arg\max_{\mathbf{x}} \, p(\mathbf{r} \mid \mathbf{x}),
\end{equation}
which is equivalent to a nonlinear \gls{wls} problem given by
\begin{equation}
\hat{\mathbf{x}} 
= \arg\min_{\mathbf{x}} 
\left(\mathbf{r} - \boldsymbol{\mu}(\mathbf{x})\right)^{\mathsf{T}}
\mathbf{C}^{-1}
\left(\mathbf{r} - \boldsymbol{\mu}(\mathbf{x})\right).
\end{equation}
The resulting optimization problem is nonlinear and can be solved using iterative optimization techniques, such as the Gauss--Newton method. The computational complexity scales with the number of \glspl{bs} and is dominated by the inversion of the covariance matrix.

\section{Simulation Setup}

\subsection{Path Loss Parameters}

To ensure realistic deployment conditions, we choose the path-loss model parameters according to the 3GPP specifications for \gls{inf} and \gls{uma} deployments \cite{3gpp.38.901, 7504435, 7109864}.
Assuming unity antenna gains, i.e., $0$ $\mathrm{dBi}$, for all antenna elements, a reference distance of $1$ $\mathrm{m}$, and a carrier frequency measured in $\mathrm{GHz}$, and using \eqref{amp}, the path loss can be expressed as 
\begin{equation}\label{eq:path_loss_dB}
    \mathrm{PL}_i \ [\mathrm{dB}] = 32 + 10\gamma \log_{10}(d_i) + 20\log_{10}(f_c),
\end{equation}
where $\gamma = 2.5$ and $\gamma = 2.8$ for the \gls{inf} and \gls{uma} scenarios, respectively.

\subsection{Network Configuration}

\begin{figure}
    \centering
    \includegraphics[width=0.725\linewidth]{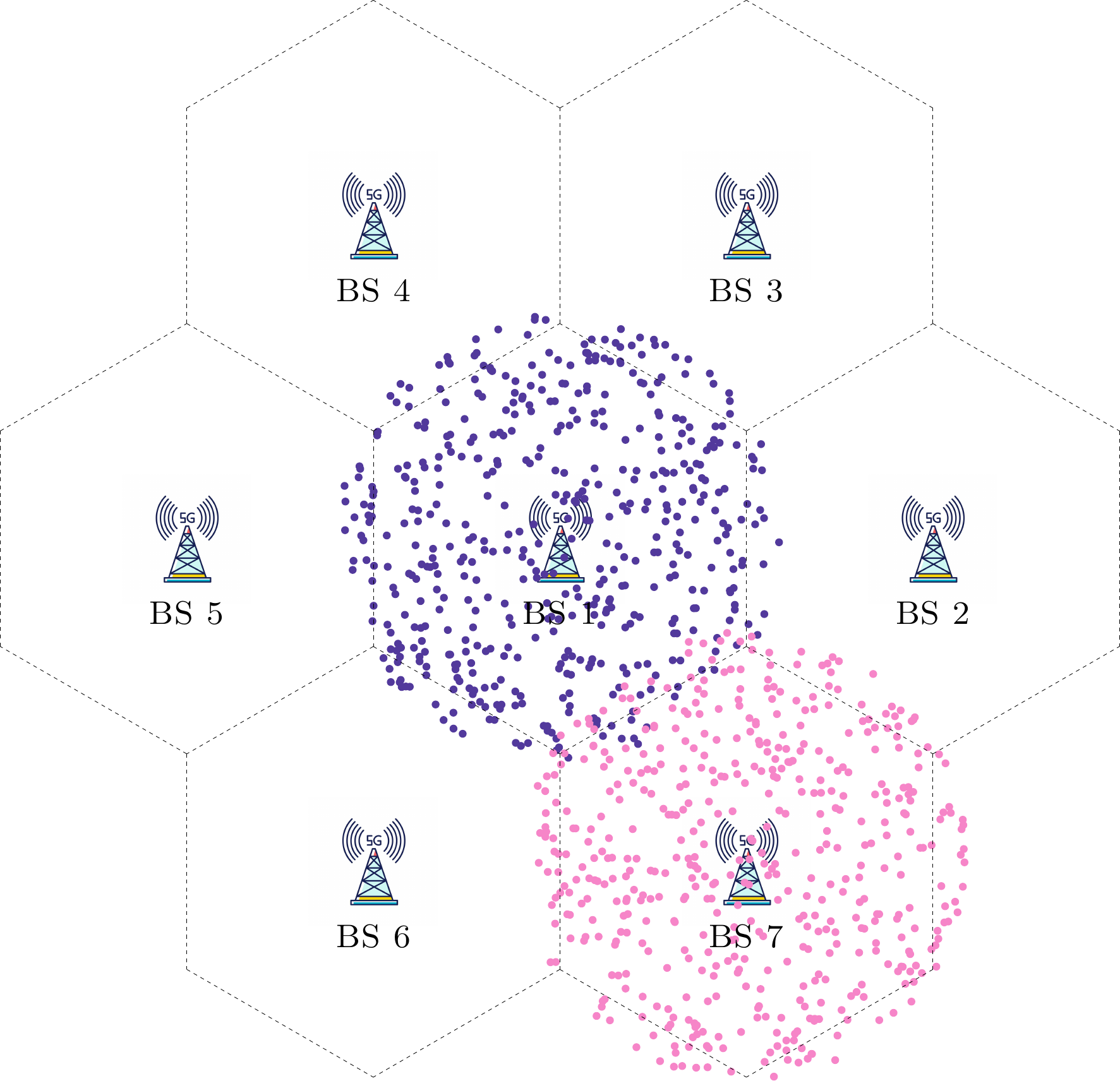}    
    \caption{Cellular layout of a 5G network with seven \glspl{bs}; the \gls{mt} is uniformly distributed within the coverage area of the reference cell, i.e., cell~1 and, for comparison, within cell~7.}
    \label{fig:default_network}
\end{figure}

We consider a cellular network comprising seven \glspl{bs}, as shown in Fig.~\ref{fig:default_network}. The \glspl{bs} are deployed in a hexagonal grid layout, which is typical for cellular networks \cite{3gpp.38.901}. Unless otherwise specified, the \gls{mt} position is assumed to be uniformly distributed over the coverage region of cell $1$. Both the \glspl{bs} and the \gls{mt} are equipped with planar phased-array antennas composed of ideal isotropic radiating elements. The total antenna gain is given as 
\begin{equation}
    G \ [\mathrm{dBi}] = G_e + G_{\mathrm{BF}},
    \label{eq:bs_gain}
\end{equation}
where $G_e$ is the gain of a single antenna element, and $G_{\mathrm{BF}}$ represents the beamforming gain \cite{10599416}. For an array with $N_e$ antenna elements, the beamforming gain can be expressed as 
\begin{equation}\label{eq:beamforming_gain}
    G_{\mathrm{BF}} \ [\mathrm{dB}] = 10\log_{10}(N_e).
\end{equation}

Considering thermal noise, the received \gls{snr} at the \gls{mt} from the $i$-th \gls{bs} can be calculated as
\begin{equation}
\begin{split}
\mathrm{SNR}_i \ [\mathrm{dB}] =  P_{\mathrm{BS}_i} & + G_{\mathrm{BS}_i}  + G_{\mathrm{MT}} - \mathrm{PL}_i \\& -  \left( N_0 + 10\log_{10}(B) \right)
\end{split}
\end{equation}

For simplicity, in the simulations of this work, identical correlation behavior is assumed across all \gls{bs} links, i.e., $\rho_{\tau_i \phi_i} = \rho$, for all $i$. The default network parameters are listed in Table~\ref{tab:simulation_parameters} and are used unless we vary a single parameter, in which case all others remain fixed at the values in Table~\ref{tab:simulation_parameters}.

\begin{table}
\centering
\begin{adjustbox}{max width=0.97\columnwidth}
\begin{tabular}{l c c}
    \toprule
    Parameter & UMa scenario & InF scenario \\
    \midrule
    Network area & $\approx 1,770,000$ $\mathrm{m}$\textsuperscript{2} & $\approx 18,000$ $\mathrm{m}$\textsuperscript{2} \\
    Number of involved BSs, $N$ & $7$ & $7$  \\
    Inter-BS distance & $500$ $\mathrm{m}$ & $50$ $\mathrm{m}$ \\
    Carrier frequency, $f_c$ & $3.75$ $\mathrm{GHz}$ & $26$ $\mathrm{GHz}$ \\
    Bandwidth, $B$ & $100$ $\mathrm{MHz}$ & $400$ $\mathrm{MHz}$ \\
    BS transmit power, $P_{\mathrm{BS}}$ & $30$ $\mathrm{dBm}$ & $24$ $\mathrm{dBm}$ \\
    BS antenna gain, $G_{\mathrm{BS}}$ & $6$ $\mathrm{dBi}$ & $12$ $\mathrm{dBi}$ \\
    MT antenna gain, $G_{\mathrm{MT}}$ & $6$ $\mathrm{dBi}$ & $12$ $\mathrm{dBi}$ \\
    BS antenna configuration & $2 \times 2$ & $4 \times 4$ \\
    MT antenna configuration & $2 \times 2$ & $4 \times 4$ \\
    Noise PSD at $293$ $\mathrm{K}$, $N_0$ & $-174$ $\mathrm{dBm/Hz}$ &  $-174$ $\mathrm{dBm/Hz}$ \\
    \bottomrule
\end{tabular}
\end{adjustbox}
\caption{5G network parameters for InF and UMa scenarios.}
\label{tab:simulation_parameters}
\end{table}

\begin{table*}[!t]
\vspace{0.09in}
\caption{Positioning RMSE in InF and UMa scenarios from Fig.~\ref{fig:rho_effect}.}
\label{tab:gain}
\centering
\begin{adjustbox}{max width=1.0\textwidth}
\begin{tabular}{l ccc ccc}
    \toprule
    & \multicolumn{3}{c}{Positioning \gls{rmse} ($E$)} 
    & \multicolumn{3}{c}{Accuracy Improvement}  \\
    \cmidrule(lr){2-4} 
    \cmidrule(lr){5-7}
    & $E_{ToA}$
    & $E_{ToA+CP (\rho = 0)}$
    & $E_{ToA+CP (\rho = 0.5)}$
    & $1-\frac{E_{\text{ToA+CP}(\rho = 0)}}{E_{\text{ToA}}}$
    & $1-\frac{E_{\text{ToA+CP}(\rho = 0.5)}}{E_{\text{ToA}}}$
    & $1-\frac{E_{\text{ToA+CP}(\rho = 0.5)}}{E_{\text{ToA+CP}(\rho = 0)}}$ \\
    \midrule
    InF
    & $1.0$ cm
    & $0.2$ cm
    & $0.1$ cm
    & $\approx 84$ $\%$
    & $\approx 85$ $\%$
    & $\approx 7$ $\%$  \\
    UMa
    & $24.7$ cm
    & $4.0$ cm
    & $3.7$ cm
    & $\approx 84$ $\%$
    & $\approx 85$ $\%$
    & $\approx 8$ $\%$  \\
    \bottomrule
\end{tabular}
\end{adjustbox}
\end{table*}

\subsection{Accuracy Metrics}
The positioning accuracy is analyzed in terms of the \gls{rmse} of the position estimate. For an empirical estimate over $N_{\mathrm{mc}}$ Monte Carlo repetitions, the \gls{rmse} of positioning is computed as 
\begin{equation}
    \mathrm{RMSE} = \sqrt{\frac{1}{N_{\mathrm{mc}}} \sum_{l=1}^{N_{\mathrm{mc}}} \|\mathbf{x} - \hat{\mathbf{x}}_l\|_2^2} \quad [\mathrm{cm}].
    \label{eq:rmse}
\end{equation}

\section{Results and Discussion}
To show the importance of accounting for the correlation between \gls{cp} and \gls{toa} measurement errors, we assess the impact of the correlation coefficient $\rho$ on positioning accuracy. We then evaluate how key 5G parameters, including bandwidth, carrier frequency, transmit power, antenna configuration, and the number of \glspl{bs}, influence the positioning accuracy of this method in real 5G networks. Hereafter, the joint correlation-aware positioning method is denoted as \textit{ToA+CP}. 

\subsection{Impact of the Correlation Coefficient}

\begin{figure}[t]
\centering
\includegraphics[width=0.85\linewidth]{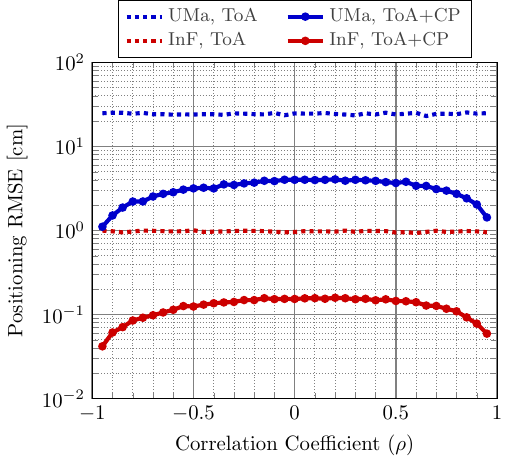}
\caption{Positioning \gls{rmse} versus correlation coefficient ($\rho$).}
\label{fig:rho_effect}
\end{figure}

Fig.~\ref{fig:rho_effect} presents the positioning \gls{rmse} as a function of the correlation coefficient $\rho$ in the \gls{uma} and \gls{inf} scenarios. As a reference, the \gls{rmse} of the conventional \gls{toa} method is also included as a horizontal line, since its performance is independent of $\rho$. The accuracy of the joint ToA+CP method depends strongly on $\rho$. The results show that the joint ToA+CP method outperforms the conventional \gls{toa} method in both \gls{uma} and \gls{inf} scenarios and that its \gls{rmse} decreases as $|\rho|$ increases. This improvement arises because the joint covariance matrix captures the statistical dependence between \gls{toa} and \gls{cp} measurement errors, which can be exploited by the maximum-likelihood estimator. From an estimation-theoretic perspective, correlation modifies the information content of the joint observations; in the considered setup, larger values of $|\rho|$ lead to a lower positioning \gls{rmse}. A slight asymmetry between positive and negative values of $\rho$ is also observed, which is attributed to the interaction between the sign of the cross-covariance terms and the nonlinear relationship between the \gls{cp} and \gls{toa} range measurements and the position estimate.

The performance gap between the InF and UMa scenarios is mainly caused by differences in their configuration parameters, as summarized in Table~\ref{tab:simulation_parameters}, such as bandwidth, inter-\gls{bs} distance, and antenna configuration. This remains true for the following results as well. Considering a moderate correlation value of $\rho = 0.5$, which may be expected in practice, Table~\ref{tab:gain} outlines the positioning errors and corresponding accuracy gains.

\begin{figure*}[t]
\centering
\includegraphics[width=0.9\linewidth]{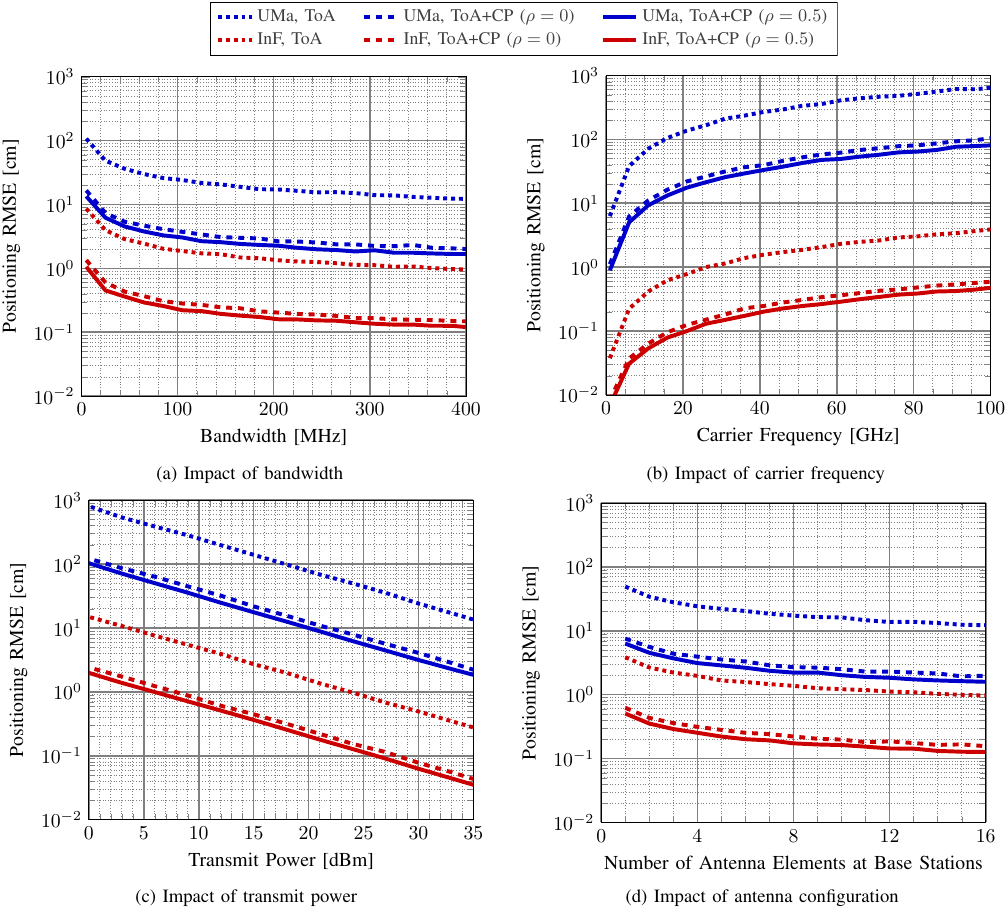}
\caption{Impact of 5G system parameters on positioning accuracy.}
\label{fig:parameter_impact}
\end{figure*}

\begin{figure*}
\centering
\includegraphics[width=0.775\linewidth]{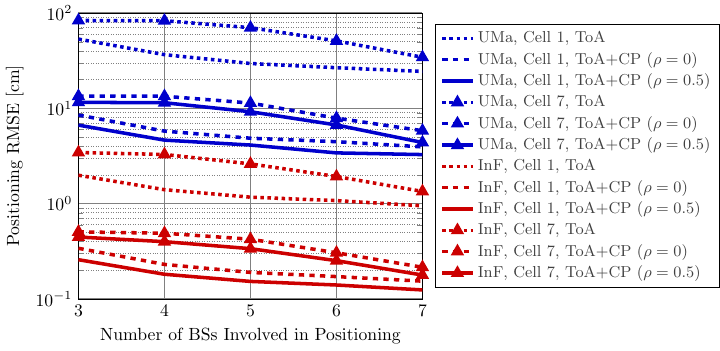}
\caption{Positioning \gls{rmse} versus the number of participating \glspl{bs} for two geometries (cell~1 and cell~7).}
\label{fig:bs_number}
\end{figure*}

\subsection{Impact of Bandwidth}
Fig.~4a illustrates the impact of bandwidth on positioning accuracy for both UMa and InF scenarios. The results show, as indicated also in \eqref{toaerror}, that increasing the bandwidth significantly improves positioning accuracy for both conventional \gls{toa} and joint ToA+CP methods. A larger bandwidth enhances the time resolution of the \gls{toa} measurements, enabling more precise delay estimation. Across the entire bandwidth range, the joint ToA+CP method consistently outperforms the \gls{toa} approach. The incorporation of carrier phase measurements further refines the range estimates, resulting in a substantial reduction in \gls{rmse}. Furthermore, the results demonstrate that accounting for correlated measurement errors, i.e., $\rho = 0.5$, yields additional accuracy gains compared to the uncorrelated case, i.e., $\rho = 0$, due to the exploitation of mutual information in both measurements. Comparing both environments, positioning errors in the InF scenario are significantly lower than those in the UMa scenario. This improvement is attributed to the shorter inter-\gls{bs} distance and higher antenna gains in InF, which lead to improved \gls{snr} and reduced measurement variance. At higher bandwidths, the joint ToA+CP method achieves centimeter-level accuracy in both InF and UMa scenarios.

\subsection{Impact of Carrier Frequency}
Fig.~4b depicts the effect of carrier frequency, varied from $1$ to $100$ $\mathrm{GHz}$. The results show that, for the setups considered, positioning accuracy degrades as the carrier frequency increases. This degradation is primarily due to increased path loss at higher frequencies, as indicated in \eqref{eq:path_loss_dB}, which reduces \gls{snr} and increases \gls{toa} estimation variance. For the joint positioning method, i.e., ToA+CP, the accuracy degradation is less pronounced. Although a higher carrier frequency increases path loss and thus reduces \gls{snr}, it simultaneously shortens the carrier wavelength, which improves phase-based ranging precision according to \eqref{toaerror}. These opposing effects partially compensate for each other, which makes the joint positioning method less sensitive to the carrier-frequency increase than the \gls{toa} method.

\subsection{Impact of Transmit Power}
Fig.~4c presents the positioning error as a function of transmit power. The results indicate a monotonic reduction in positioning error with increasing transmit power for all methods. Higher transmit power directly increases the received \gls{snr}, which reduces both \gls{toa} and \gls{cp} measurement errors according to \eqref{toaerror} and thereby lowers the positioning \gls{rmse}. As in all other cases, the results show that accounting for correlated measurement errors, i.e., $\rho = 0.5$, yields additional accuracy gains compared to the uncorrelated case, i.e., $\rho = 0$.

\subsection{Impact of Antenna Configuration}
Fig.~4d shows the influence of the number of antenna elements in the phased array of \glspl{bs}, varying from 1 to 16. Increasing the number of antenna elements improves positioning accuracy in both scenarios. This improvement is attributed to the beamforming gain described in \eqref{eq:beamforming_gain}, which enhances transmit directivity and thereby increases the \gls{snr} at the \gls{mt}. The improved \gls{snr} reduces measurement variances for both \gls{toa} and \gls{cp}, leading to lower positioning \gls{rmse}. The results demonstrate that antenna array design plays a crucial role in enabling high-accuracy positioning, especially in high-frequency deployments where path loss is severe.

\subsection{Impact of the Number of BSs Involved in Positioning}

Fig.~\ref{fig:bs_number} shows how positioning accuracy is influenced by (i) the number of \glspl{bs} included in the estimation, which are added sequentially according to their indices, and (ii) the deployment geometry. Two \gls{mt} locations are considered: (a) a user uniformly distributed within cell~$1$, representing a geometrically favorable configuration, and (b) a user uniformly distributed within cell~$7$, representing a less favorable geometry. The results indicate that positioning accuracy generally improves as more \glspl{bs} are incorporated. This improvement is attributed to enhanced geometric diversity and measurement redundancy. However, the magnitude of this improvement depends strongly on the \gls{mt} location. When the \gls{mt} is in cell~$1$, the surrounding \glspl{bs} form a favorable spatial configuration; therefore, a substantial reduction in positioning error is achieved with only a few \glspl{bs}, and adding further \glspl{bs} yields diminishing returns. In contrast, when the \gls{mt} is in cell~$7$, the initial geometry is less favorable, and incorporating additional \glspl{bs} continues to provide noticeable improvements in accuracy.

\section{Conclusions}
In this paper, we propose and study a joint positioning method based on \gls{toa} and \gls{cp} measurements in 5G networks, explicitly accounting for the correlation between their measurement errors, as both are extracted from the same received signal and may therefore exhibit correlated errors.
The accuracy of the proposed joint positioning method was evaluated as a function of key network parameters using path-loss parameters for the 3GPP-standardized \gls{inf} and \gls{uma} environments.
The joint ToA+CP positioning method consistently outperformed the conventional \gls{toa} approach.
Moreover, the results show that explicitly accounting for measurement-error correlation yields additional positioning accuracy gains in both deployment scenarios. The simulation results further indicate that, under favorable geometry and high-\gls{snr} conditions, 5G positioning can achieve centimeter-level accuracy.
For future research, we suggest conducting empirical studies to determine the magnitude and distribution of the correlation coefficient between \gls{toa} and \gls{cp} measurements, $\rho$, in practical implementations.

\bibliographystyle{IEEEtran}
\bibliography{references}

\end{document}